\pdfoutput=1
\documentclass{aa}    
\bibpunct{(}{)}{;}{a}{}{,} 

\usepackage{graphicx}
\usepackage[varg]{txfonts}
\usepackage{amssymb,amsmath,cancel}
\usepackage[dvipsnames]{xcolor}  
\usepackage{soul}  
\usepackage{multicol}  

\usepackage{ulem} 

\def\gapprox{\;\rlap{\lower 2.5pt            
 \hbox{$\sim$}}\raise 1.5pt\hbox{$>$}\;}       
\def\lapprox{\;\rlap{\lower 2.5pt            
 \hbox{$\sim$}}\raise 1.5pt\hbox{$<$}\;} 
\begin{document}
\title{A well-balanced scheme for the simulation tool-kit A-MaZe\,: implementation, tests, and first
  applications to stellar structure}
   \titlerunning{Stellar structure by A-MaZe}
   \author{M.~V.~Popov\inst{1},
          R.~Walder\inst{1},
          D.~ Folini\inst{1},
          T.~Goffrey\inst{2},
          I. ~Baraffe\inst{2,1},
          T.~Constantino\inst{2},
          C.~Geroux\inst{2},
          J. ~Pratt\inst{2,3}
          M.~Viallet\inst{2}
\and
          R. K\"appeli\inst{4}
}
   \authorrunning{M.~V.~Popov, R.~Walder, D.~ Folini et al.}
   \institute{\'{E}cole Normale Sup\'{e}rieure, Lyon, CRAL, UMR CNRS 5574, 
           Universit\'{e} de Lyon, France\\
           \email{doris.folini@ens-lyon.fr}
%
\and
	College of Engineering, Mathematics and Physical Sciences, University of Exeter, Exeter, EX4 4QL, UK
\and
 Department of Physics and Astronomy, Georgia State University, Atlanta, GA 30303, USA
\and
 Seminar of Applied Mathematics, ETH-Z\"urich, Switzerland
}
   \date{Received ... ; accepted ...}
  \abstract
{ Characterizing stellar convection in multiple dimensions is a topic
  at the forefront of stellar astrophysics. Numerical simulations are
  an essential tool for this task. We present an extension of the
  existing numerical tool-kit A-MaZe that enables such simulations of
  stratified flows in a gravitational field. The finite-volume based,
  cell-centered, and time-explicit hydrodynamics solver of A-MaZe was
  extended such that the scheme is now well-balanced in both momentum
  and energy. The algorithm maintains an initially static balance
  between gravity and pressure to machine precision. Quasi-stationary
  convection in slab-geometry preserves gas energy (internal plus
  kinetic) on average despite strong local up- and down-drafts. By
  contrast, a more standard numerical scheme is demonstrated to result
  in substantial gains of energy within a short time on purely
  numerical grounds. The test is further used to point out the role of
  dimensionality, viscosity, and Rayleigh number for compressible
  convection. Applications to a young sun in 2D and 3D, covering a
  part of the inner radiative zone as well as the outer convective
  zone, demonstrate that the scheme meets its initial design
  goal. Comparison with results obtained for a physically identical
  setup with a time-implicit code show qualitative agreement. }


   \maketitle
\section{Introduction}
\label{Sec:Introduction}
In a wide range of astrophysical objects, the balance between
gravitation and gas pressure is a key element, on top of which
additional physics may take place. Ascertaining the robustness of
corresponding numerical results by applying different codes to the
same physical problem motivates this paper about the adaptation of
the A-MaZe tool-kit~\citep{2000ASPC..204..281W, 2003ASPC..288..433F,
  2013A&A...558A.133M} so that it can tackle gravitationally
stratified flows.

Stars and planets are prominent astrophysical examples harboring such
flows. Although quasi-static in a global sense, a wealth of dynamics
takes place. Energy transport via convection is often essential to
maintain a globally quasi-static state. This is notably the case
for the outer parts of low mass stars and the inner regions of high
mass stars, where energy transport via radiation is not efficient
enough. Multidimensional simulations of such transport processes
contribute to the 321D link for stellar
modeling~\citep[e.g.][]{2015ApJ...809...30A}, i.e., the effort to
improve one-dimensional stellar evolution models via simulating short
episodes in 2D and 3D. Such studies motivate the overall project this
study is embedded in~\citep[see e.g.][]{2016A&A...588A..85G,
  2017A&A...604A.125P, 2017ApJ...845L...6B}.  Other examples exist
where the pressure-gravity balance plays a crucial role and associated
numerical challenges resemble the ones we are interested in here. We
mention the vertical structure of accretion disks and supernova
explosions, notably the moment just before the onset of the collapse
and then again when the prompt shock
stalls~\citep{2013ApJ...778L...7C, 2016ApJ...833..124M,
  2017MNRAS.472..491M}.

From a numerical point of view, the above astrophysical problems are
challenging because the gravity and pressure forces - both typically large,
but of opposite sign  - may not cancel in
their discretized form. This non-cancellation may severely impact the
solution. A number of remedies have been suggested~\citep[for a short
  overview, see e.g.][]{2016A&A...587A..94K}. Here we focus on so
called well-balanced schemes, i.e., schemes which are designed to
exactly maintain a discrete equivalent of the underlying stationary
state. Initially put forward by~\citet{cargo-leroux:94} and~\citet{Greenberg1996}, numerous
concrete forms of well-balanced schemes meanwhile
exist~\citep{1998JCoPh.146..346L, Noelle2009, 2009JCoPh.228.6682W,
  Xing2013, 2014JCoPh.259..199K, 2016A&A...587A..94K,
  2015SIAMJSC..37..B382C, 2016HypAcchen2016...B, 2016IJNMF..81..104D,
  Touma2016}. Major differences among them include the allowed
equation of state (EoS), roughly speaking 'simple' or 'complicated',
whether the EoS explicitly enters the well-balanced reconstruction or
not, the required knowledge about the stationary state, from
'analytical form' to 'none at all', whether the method is designed for
shallow water problems or the full Euler or Navier-Stokes equations,
and whether the scheme relies on velocities being (and remaining)
identically to zero or not.

The goal of this paper is to document the implementation of a
well-balanced scheme into the A-MaZe simulation tool-kit, to
demonstrate its performance with a series of tests, to illustrate the
behavior of a not well-balanced scheme for the same tests, and to draw
attention to selected physical aspects of the tests.

As our ultimate interest is with stellar convection, we want a
well-balanced scheme that is applicable to the full Euler or Navier-Stokes
equations, works for both flows in external gravitational fields and for
self-gravitating flows, can cope with an arbitrary EoS, does not rely
on a priori knowledge of the stationary state, and can accommodate
non-zero velocities. The scheme should be simple enough such that it
can be easily accommodated in an existing code and, in the future, can
be combined with adaptive meshes and general curvilinear grids. With these
considerations in mind, we decided for the approach
by~\citet{2016A&A...587A..94K} (KM16 in the following) as a starting
point for our own work, whose treatment of the energy source term we
adapt.

The paper is structured as follows. We present the overall problem
along with our algorithm, as part of the simulation tool-kit A-MaZe
and with particular focus on the well-balanced scheme, in
Sect.~\ref{Sec:Algorithm_Code}. A first set of test cases, static
configurations and convective slabs, are examined in
Sect.~\ref{Sec:Tests}. The performance of the algorithm for a test
case closer to our finally envisaged applications, featuring in
particular a general EoS that comes in the form of a
look-up table, is demonstrated in Sect.~\ref{Sec:YoungSun}. Discussion
follows in Sect.~\ref{Sec:Discussion}, summary and conclusions in
Sect.~\ref{Sec:Conclusion}.
\section{Equations and methods}
\label{Sec:Algorithm_Code}
We look for numerical solutions of the compressible
Navier-Stokes equations in the presence of a gravitational
field, as detailed in Sect.~\ref{Sec:Algorithm_Code_Equations}.
Essential aspects of A-MaZe, our hosting numerical tool-kit, are
summarized in Sect.~\ref{Sec:Algorithm_Code_Implementation}. The
well-balanced extension of A-MaZe is detailed in
Sect.~\ref{Sec:Algorithm_Code_WBScheme}
\subsection{Navier-Stokes Equations  with Gravity}
\label{Sec:Algorithm_Code_Equations}
The Navier-Stokes equations for a thermally conductive and
compressible medium in the presence of a gravitational potential $\phi$ can be
written as a system of balance laws in the form
\begin{equation}
\label{Eq:Hydro1}
\frac{\partial\rho}{\partial t}+ \nabla(\rho\,{\bf u}) = 0,
\end{equation}
\begin{equation}
\label{Eq:Hydro2}
\frac{\partial(\rho\,{\bf u})}{\partial t} + \nabla(\rho\,{\bf u}\otimes{\bf u}) + \nabla p-\nabla\tau = -\rho\,{\nabla\phi}\,,
\end{equation}
\begin{equation}
\label{Eq:Hydro3}
\frac{\partial E}{\partial t} + \nabla \left( {\bf u} ( E + p) \right)-\nabla\left(K\nabla T\right)-\nabla
(\tau\,{\bf u}) = -\rho\,{\bf u}\ {\bf \nabla}\phi\,,
\end{equation}
with 
\begin{equation}
\label{Etotal}
E=\rho e+\frac{\rho{\bf u}^2}2\,
\end{equation}
the gas energy density, $\rho$ the density, $\bf{u}$ the velocity
vector, $p$ the gas pressure, $e$ the specific internal energy, $K$
the (potentially non-linear) heat-transfer coefficient, and $T$ the
temperature. The components of the dynamic viscous stress tensor, denoted $\tau$, are defined as
\begin{equation}
\label{Visc}
\tau_{ij}=\mu\left(\frac{\partial u_i}{\partial x_j}+\frac{\partial
  u_j}{\partial x_i}-\frac23\delta_{ij}\frac{\partial u_k}{\partial
  x_k}\right),
\end{equation}
where $\mu$ is the dynamic viscosity (volume or bulk viscosity is
neglected).  The EoS as well as $\mu$ and $K$, are problem specific
and, consequently, are further specified along with the test cases in
Sects.~\ref{Sec:Tests} and~\ref{Sec:YoungSun}.

Throughout the paper we neglect changes of gravity due to matter
re-distribution. We assume a problem specific but time-constant
gravitational potential $\phi$ with associated free-fall acceleration
vector $\bf{g}$
\begin{equation}
\label{PHI}
{\bf g}=-\nabla{\phi}.
\end{equation}
Hydrostatic equilibrium is therefore defined by 
\begin{equation}
\label{eq:hydrostat}
\nabla p=-\rho\nabla\phi.
\end{equation}
The gravitational energy density $E_{g}$ of the gas is given by
\begin{equation}
\label{Eq:EGrav}
E_{g}=\frac{1}{2}\rho\phi.
\end{equation}
Two characteristic time scales are the sound crossing time
\begin{equation}
\label{sound_cross}
\tau_{\mbox{s}} = \int_{x_0}^{x_1}\frac{dx}{c_s(x)}
\end{equation}
and the convective
turnover time $\tau_{c}$
\begin{equation}
  \tau_{\mbox{c}} =  \int_{x_0}^{x_1}\, \frac{dx}{\left|u_{x}(x)
    \right|} = \int_{x_0}^{x_1}\,dx\, \frac{1}{M(x)} \cdot
  \frac{1}{c_s(x)} \approx \frac{\tau_{\mbox{s}}}{\overline{M}}.
\label{Eq:ConvectiveTurnOver_Time_Def}
\end{equation}
The integral extends over a characteristic length scale, e.g a scale
height or the depth of convection zone (here in x-direction). The
local (at distance $x$) sound speed, Mach number, and flow velocity in
direction of ${\bf g}$ are denoted by $c_s(x)$, $M(x)$, and
$u_{x}(x)$, respectively. $\bar{M}$ is the depth-averaged Mach number.
In many astrophysically relevant cases, the convective motion is very
subsonic and thus $\tau_{\mbox{c}} >> \tau_{\mbox{s}}$. For instance,
in their study of a young sun in 2D with $M \le 0.05$,
\citet{2016A&A...593A.121P} needed an integration time of several
hundred $\tau_{\mbox{c}}$, translating into several thousand
$\tau_{\mbox{s}}$, for robust statistics. This sets the time scale of
interest, over which Eqs.~\ref{Eq:Hydro1} to~\ref{Eq:Hydro3} have to
be integrated numerically.
\subsection{Numerical tool-kit A-MaZe}
\label{Sec:Algorithm_Code_Implementation}
We build on the numerical tool-kit A-MaZe~\citep{2000ASPC..204..281W,
  2003ASPC..288..433F, 2013A&A...558A.133M}, a collection of adaptive
mesh \citep{1984JCoPh..53..484B, 1989JCoPh..82...64B,
  2003ASPC..288..433F} multi-scale, multi-physics codes and analysis
tools to support simulations of astrophysical objects. A-MaZe has been
applied to a range of problems, including accretion and blasts in
novas \citep{2008A&A...484L...9W}, full scale simulations of X-ray
binaries \citep{2014ASPC..488..141W}, colliding winds and emitted
spectra \citep{1993A&A...278..209N, 1999IAUS..193..352F,
  2000Ap&SS.274..189F}, particle acceleration in relativistic magnetic
reconnection \citep{2014A&A...570A.111M, 2014A&A...570A.112M},
investigation of supersonic turbulence \citep{2006A&A...459....1F,
  2014A&A...562A.112F}, and the dynamics of circum-stellar material
\citep{2004A&A...414..559F, 2013A&A...559A..69G}.

Until now, A-MaZe lacked a proper treatment of a static balance between
gravitational and pressure forces. The present paper remedies this
shortcoming by appropriately augmenting (see
Sect.~\ref{Sec:Algorithm_Code_WBScheme}) the hydrodynamic solver.  Key
characteristics of the latter, as far as relevant for and used in this
paper, are summarized in the following.

Eqs.~\ref{Eq:Hydro1} to \ref{Eq:Hydro3} are solved using a finite
volume discretization on the basis of mapped grids
\citep{2008SIAMR..50..723C} for general curvi-linear coordinates. A
regular Cartesian mesh (the computational mesh) is mapped to the
desired mesh in physical space (the physical mesh). Within this paper,
physical meshes are truly Cartesian (in 1D, 2D, and 3D), 2D
axi-symmetric, or 3D spherical shell wedges. Other mappings have
  not yet been implemented, although the code infrastructure to hold
  them is in place. As the concrete mapping functions depend on the
specific application, they are formulated in the context of the latter
in Sect.~\ref{Sec:Tests}.  Advective fluxes through the cell faces are
computed with the help of a Riemann-solver, which feeds on data at
cell interfaces (left and right state, at the geometrical center of
the interface) obtained by standard (limited) reconstruction
techniques. Within the context of mapped grids, advective as well as
diffusive fluxes can be computed on the computational mesh, with the
physical mesh entering only via cell surface areas and cell
volumes. This procedure avoids geometrical source terms~\citep[see
  e.g.][for a more detailed discussion]{2012A&A...544A..47K}.

The semi-discretized three-dimensional version of Eqs.~\ref{Eq:Hydro1} to~\ref{Eq:Hydro3} is
\begin{eqnarray}
\label{Eq:MOL}
\frac{\partial{\bf U}_{i,j,k}}{\partial t} 
& + & \frac{S_{i+1/2,j,k}{\bf F}_{i+1/2,j,k}-S_{i-1/2,j,k}{\bf F}_{i-1/2,j,k}}{V_{i,j,k}} + \nonumber \\
&   & \frac{S_{i,j+1/2,k}{\bf G}_{i,j+1/2,k}-S_{i,j-1/2,k}{\bf G}_{i,j-1/2,k}}{V_{i,j,k}} + \nonumber \\
&   & \frac{S_{i,j,k+1/2}{\bf H}_{i,j,k+1/2}-S_{i,j,k-1/2}{\bf H}_{i,j,k-1/2}}{V_{i,j,k}}={\bf \Psi}_{i,j,k}.
\end{eqnarray}
${\bf U}_{i,j,k}=\left(\rho,\rho u_{x},\rho u_{y},\rho
u_{z},E\right)^{T}_{i,j,k}$ is the vector of conserved quantities at
cell centers $(i,j,k) \in (1,\ldots, N_{x}, 1,\ldots, N_{y}, 1,\ldots, N_{z})$,
with $N_{x}, N_{y}$, and $N_{z}$ the number of cells in $x-$, $y-$, and
$z-$direction of computational space. Half indices denote cell
faces. ${\bf F}_{i\pm1/2,j,k}$, ${\bf G}_{i,j\pm1/2,k}$, and ${\bf
  H}_{i,j,k\pm1/2}$ denote the fluxes through the cell faces. Note
that they contain advective and diffusive terms. Physical space enters
only via the cell surfaces (terms $S_{i\pm1/2,j,k}$) and cell volumes
(terms $V_{i,j,k}$), both evaluated in physical space, as well as via
the source term ${\bf \Psi}_{i,j,k}$, also evaluated in physical
space.

Time integration of Eq.~\ref{Eq:MOL} in this paper is done with a
first order Runge-Kutta method, i.e., a forward Euler method,
although A-MaZe also offers strong stability preserving (SSP) higher
order integration schemes~\citep{1988JCoPh..77..439S,
  2001SIAMR..43...89G}. We use different Riemann solvers, notably the
approximate HLLC~\citep{1994ShWav...4...25T}, as well as the exact
solver by~\citet{1985JCoPh..59..264C}. Likewise, we used both minmod
and van Leer limiters together with second order reconstruction.
Results seem overall robust to the particular choice of integrator and
limiter, but no detailed analysis in this respect was done. 
  Diffusive fluxes are approximated by a dimensionally split second
  order central finite difference discretization in space, i.e., each
  spatial direction is treated independently of the others.  In the
following, we refer to the above approach as the standard scheme.

A variant better suited for situations with a gravitational field, in
the following referred to as well-balanced scheme, is described in
Sect.~\ref{Sec:Algorithm_Code_WBScheme}. The difference between the
two schemes is small in terms of code: it concerns only the
reconstruction of pressure at cell interfaces and the discretization
of the source term on the right hand side of Eq.~\ref{Eq:Hydro3}.
\subsection{The well-balanced algorithm}
\label{Sec:Algorithm_Code_WBScheme}
With the gravitational potential related source terms for momentum and
energy, $S^{M}$ and $S^{E}$ on the right hand side of
Eqs.~\ref{Eq:Hydro2} and~\ref{Eq:Hydro3}, respectively, care must be
taken to avoid associated numerical sources or sinks of momentum and
energy.
\subsubsection{Momentum balance}
\label{SubSec:WB_MomentumBalance}
The momentum equation, Eq.~\ref{Eq:Hydro2}, includes the hydrostatic
balance equation, Eq.~\ref{eq:hydrostat}, to which it reduces in the
case of zero velocities. Unless the numerical scheme respects this
balance to machine precision, the momentum equation will suffer from
spurious gains / losses of momentum associated with $S^{M}$.

We adopt the scheme of KM16, the key aspects of which we summarize
here.  The overarching idea is to arrive at a reconstructed pressure
at cell interfaces that is a) identical on the left and right side of
the interface and b) respects a discrete version of the hydrostatic
equilibrium equation, Eq.~\ref{eq:hydrostat}.  Condition a) ascertains
that the Riemann solver is not fed any spurious pressure jumps that
would be translated into waves. Condition b) ascertains that the
pressure gradient across a cell - the pressure contribution to the
flux differencing term - matches the source term of the discrete
hydrostatic equilibrium. As is stressed in KM16, the equilibrium is a
mechanical one, no assumption is made on a thermal equilibrium, i.e.,
no explicit temperature or entropy profile has to be assumed and the
hydrostatic equilibrium may be physically unstable to
convection. Perturbations on top of the hydrostatic equilibrium
pressure are addressed by decomposing the total pressure for the
reconstruction as $p=p^0+p^1$, with the hydrostatic part $p^0$ and the
perturbation part $p^1$. The well-balanced reconstruction of $p^{0}$
is second order. For $p^{1}$ and all other variables, any higher order
reconstruction may be used. We use second order in this paper with a
minmod limiter.

In 1D (arbitrarily in x-direction) the relevant formulas read as
follows.  The gravitational source term $-\rho\,{\nabla\phi}$ on the
right hand side of Eq.~\ref{Eq:Hydro2} is discretized as a centered
difference,
\begin{equation}
\label{Sm_def}
S^{M}_i = - \rho_i\frac{\phi_{i+1}-\phi_{i-1}}{x_{i+1}-x_{i-1}}.
\end{equation}
The well-balanced reconstruction of the equilibrium part $p^0$ of the
pressure of cell $i$ is given by
\begin{equation}
\label{int_res}
\begin{aligned}
p^0_i(x_{i-1/2})=p_i+\rho_i\frac{\phi_i-\phi_{i-1}}{x_i-x_{i-1}}\left(x_i-x_{i-1/2}\right),\\
p^0_i(x_{i+1/2})=p_i-\rho_i\frac{\phi_{i+1}-\phi_i}{x_{i+1}-x_i}\left(x_{i+1/2}-x_i\right).
\end{aligned}
\end{equation}
With $p^0_i(x_{i+1/2}) = p^0_{i+1}(x_{i+1/2})$ (see condition a)
above), Eq.~\ref{int_res} leads to the discrete, spatially
second-order accurate form of the hydrostatic equilibrium equation,
Eq.~\ref{eq:hydrostat},
\begin{equation}
\label{Eq_discrete}
\frac{p_{i+1}-p_i}{x_{i+1}-x_i}=-\frac{\rho_i+\rho_{i+1}}{2}
\frac{\phi_{i+1}-\phi_i}{x_{i+1}-x_i}.
\end{equation}
The reconstruction of the perturbation part $p^1$ of the pressure
(which can be of any order, see KM16) takes as input cell
centered values
\begin{equation}
\label{p_pertub}
\begin{aligned}
p^1_i(x_{i-1})=p_{i-1}-p^0_i(x_{i-1}),\\
p^1_i(x_{i+1})=p_{i+1}-p^0_i(x_{i+1}),
\end{aligned}
\end{equation}
with
\begin{equation}
\label{int_res_extra}
\begin{aligned}
p^0_i(x_{i-1})=p_i+\frac{\rho_{i-1}+\rho_i}2\left(\phi_i-\phi_{i-1}\right),\\
p^0_i(x_{i+1})=p_i-\frac{\rho_i+\rho_{i+1}}2\left(\phi_{i+1}-\phi_i\right).
\end{aligned}
\end{equation}
\subsubsection{Energy Balance}
\label{SubSec:WB_EnergyBalance}
The source term $S^E$ in the energy equation, Eq.~\ref{Eq:Hydro3},
mediates between the gravitational potential energy $E_{g}$ of the gas
and its internal plus kinetic energy $E$: as mass is advected along
the gravitational field, energy is transferred from $E_{g}$ to $E$ and
vice versa. A numerical pitfall then opens: with $\phi$ constant in
time there is an unlimited reservoir of gravitational energy, thus
arbitrary amounts of gas energy may be gained or lost on numerical
grounds if $S^E$ is not accurately computed.  The issue is particularly
relevant for quasi-stationary convection, where $E_{g}$ and $E$ are
constant when integrated over the domain of interest, although there is
a steady exchange between the two on local scales: in down-flows energy
is transferred from $E_{g}$ to $E$ whereas the opposite is true in
regions of up-flow. The numerical scheme must respect the global
balance despite the steady local exchange.

KM16 use centered differences for the source term in
Eq.~\ref{Eq:Hydro3}.  As we illustrate in Sect.~\ref{Sec:Tests}, this
choice leads to a systematic increase of $E$ with time when simulating
quasi-stationary convection. This although there is no net motion of
mass in the direction parallel to the gravitational force.

We use an alternative formulation, expressing $S^E$ in terms of
discrete mass fluxes $F^M$, $G^M$, and $K^M$ as
\begin{eqnarray}
\label{Eq:Energy_Source_Term}
S^E_{i,j,k} & = & 0.5 (F^M_{i+1/2,j,k} g_{i+1/2,j,k} + F^M_{i-1/2,j,k} g_{i-1/2,j,k} + \nonumber \\ 
           &   & G^M_{i,j+1/2,k} g_{i,j+1/2,k} + G^M_{i,j-1/2,k} g_{i,j-1/2,k} + \nonumber \\
           &   &K^M_{i,j,k+1/2} g_{i,j,k+1/2} + K^M_{i,j,k-1/2} g_{i,j,k-1/2} ).
\end{eqnarray}
Here, $g_{i+1/2,j,k} = (\phi_{i+1,j,k} - \phi_{i,j,k}) / (x_{i+1,j,k}
- x_{i,j,k})$ etc.  The expression for $S^E$ can be motivated in two
ways. From a physical point of view, one may argue with the connection
between the mass flux, projected onto $\nabla \phi$, and the local
exchange between $E_{g}$ and $E$. The above equation ascertains that
the domain average of $S^{E}_{i,j,k}$ is zero - there is no domain
averaged net exchange between $E$ and $E_{g}$ - unless there is some
domain averaged net mass flux parallel to $\nabla \phi$. Another way
to obtain Eq.~\ref{Eq:Energy_Source_Term} is to start from a
conservative discretization for the total energy $E_{tot} = E +
E_{g}$, instead of using a source term in the energy equation. This is
done, for example, in~\citet{2013NewA...19...48J}. Starting from their
equations for the total energy (Eq.~9), for the energy flux ${\bf
  F}_{g}$ due to gravity (Eq.~14), and for the update of the energy
(Eq.~16), noting in addition that in our case the gravitational
potential $\phi$ is time constant and the gravitational potential
energy is given by Eq.~\ref{Eq:EGrav}, and using for the components of
${\bf F}_{g}$ the discrete expressions $F^M_{i+1/2,j,k} g_{i+1/2,j,k}$
etc. one obtains Eq.~\ref{Eq:Energy_Source_Term}.

For an individual time step, we expect the difference between our
formulation and a centered difference formulation for $S^E$ to be
small. This is because the mass fluxes at cell interfaces must closely
resemble the mass fluxes evaluated at cell centers, the difference
arising in essence from the reconstruction (from cell center to cell
interface). An analytical error estimate is, however, not trivial as
we use the fluxes from the (approximate) Riemann solver. The
difference between the two approaches is, however, systematic and thus
cumulative over many time steps, at least for a stratified medium in
the presence of a fixed gravitational field. The numerical results in
Sect.~\ref{Sec:Tests} will further underpin this point.
\section{Simulating Convection: Basic Tests}
\label{Sec:Tests}
Our goal here is twofold: we want to demonstrate that the
well-balanced scheme lives up to expectations and we want to
illustrate the consequences of using a standard scheme.
Sect.~\ref{Sec:Tests_StaticAtmospheres} focuses on static (zero
velocity), marginally stably stratified situations with polytropic EoS
in different geometries. Sect.~\ref{Sec:Tests_Convection} addresses
quasi-stationary convection, from 1D to 3D, for an ideal gas EoS.
The description of each test follows the same basic scheme: sketch of
the physical problem (equations, boundary conditions, analytical
solution where possible), information on the numerical domain and its
discretization, details on problem initialization and numerical
boundary conditions, and, finally, the results of the test.
\subsection{Static stratified layers}
\label{Sec:Tests_StaticAtmospheres}
Tests in this section consist of a stratified medium at
rest. Integration in time of such a configuration may or may not
preserve its initial, static state, depending on whether the
integration scheme is well-balanced or not.
\subsubsection{Plane-parallel 1D slab}
\label{Sec:Tests_QuasiStatic_Slab}
\begin{figure}[tbp]
\centering
\includegraphics[width=0.9\linewidth]{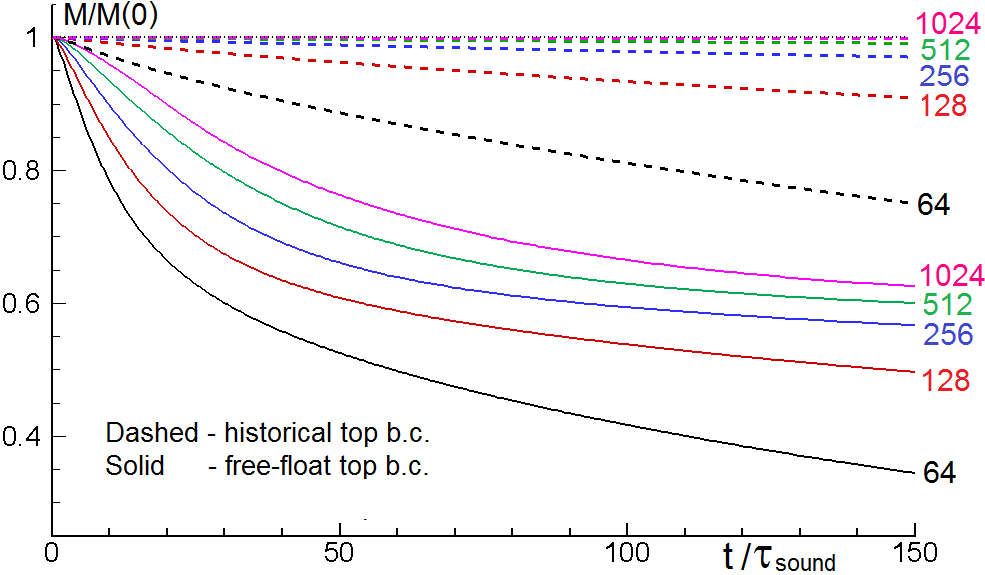}
\caption{Total mass in 1D hydrostatic slab, normalized by the initial
  mass ($M(t)/M(t=0)$, y-axis), as function of time (x-axis, in
  $\tau_{\mathrm{s}}$), integrated with the standard scheme, using
  either historical (dashed lines) or free-float (solid lines)
  boundary conditions and different meshes (colors; number of grid
  cells).  The well-balanced scheme maintains $M(t)/M(t=0) = 1$ to
  machine precision in all cases.}
\label{Fig:Mass_leaking}
\end{figure}
We repeat one of the tests from KM16, their section 3.1: a hydrostatic
1D plane-parallel atmosphere with gravitational potential
$\phi(x)=g\,x$ and the EoS of a mono-atomic ideal, isentropic gas,
$p\,(\rho,s)=e^{s/c_{v}}\rho^{\gamma}$ with
$s=s_0=R_{gas}/(\gamma-1)\ln(p_0/\rho_0^{\gamma})$.  We use $g=1$,
$\gamma=5/3$, and $R_{gas}= 1$.  The analytical solution of
Eq.~\ref{eq:hydrostat} under these conditions is given by
\begin{equation}
\label{Eq1D_setup1}
\rho(x)=\left(\rho_0^{\gamma-1}-e^{-s/c_{v}}\frac{\gamma-1}{\gamma}g\,x\right)^{1/(\gamma-1)}.
\end{equation}
For our numerical domain $x\in[0,2]$ pressure scale heights
$h_p(x)=-p/(dp/dx)$ range from $h_p(0)=1$ to $h_p(2)\sim0.578$. Meshes
range from 64 cells to 1024 cells. Thermal and viscous diffusion terms
are set to zero in the numerical solution.

As our goal is to test whether our well-balanced scheme can maintain
to machine precision a configuration that is initially stratified and
at rest, we must ascertain that the initialization satisfies
Eq.~\ref{Eq_discrete}, the discrete counterpart of
Eq.~\ref{eq:hydrostat}. To this end, we anchor our discrete solution at the
lower domain boundary, where we set $\rho_0=1$ and $p_0=1$ (implying
$s_0=0$). We then use a Newton-Raphson method to integrate
Eq.~\ref{Eq_discrete} together with the above EoS towards the upper
domain boundary.  The velocity is set to zero everywhere.

Two sets of numerical boundary conditions are used, employing again
the Newton-Raphson method to integrate the analytical solution
  (given here by Eqs.~\ref{Eq_discrete} and~\ref{Eq1D_setup1}) from
  the last within domain cell to fill the ghost cells before each time
step. The bottom boundary is reflecting in both cases. The top
boundary is set to either of the following:
\begin{itemize}
\item{{\bf Free-float\,}: the current density and pressure profile is extrapolated, velocities are copied from within the domain.}
\item{{\bf Historical\,}: the initial density and pressure profile is extrapolated, velocities are set to zero.}
\end{itemize}
The free-float boundary conditions correspond to boundary conditions
as used in KM16. The historical boundary conditions are a variant
thereof, designed to draw the boundaries in each time step to
the initial state.

Using the standard scheme, average velocities are around $10^{-6}$ to
$10^{-4}$ after only $2\, \tau_{\mathrm{s}}$. The sound speed, for
comparison, is around one. KM16 give a more detailed error analysis of
these non-zero velocities. Here we want to draw attention to a
follow up effect:
a net transfer of mass towards the upper boundary and a decrease of
the mass $M(t)$ of the slab with time, as illustrated in
Fig.~\ref{Fig:Mass_leaking}. As can be seen, the mass loss depends on
the discretization (more severe for coarser grids) and on the boundary
conditions (more severe for free-flow). The well-balanced scheme
maintains zero velocities and $M(t)/M(0)=1$ to machine precision for
both boundary conditions, even for the most coarse discretization of
64 cells.

The test demonstrates not only that the well-balanced scheme works as
expected, but also that it is necessary: the solution suffers from
severe numerical artifacts if the standard scheme is used.
\subsubsection{Lane-Emden polytrope in 2D and 3D}
\label{Sec:Tests_QuasiStatic_Polytrope_AxiSym}
\begin{figure}[tbp]
\centerline{\includegraphics[width=0.95\linewidth]{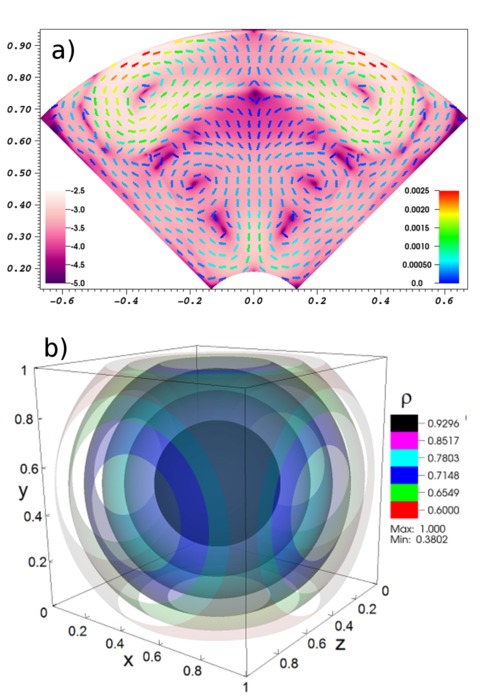}}
\caption{Lane-Emden test case, preservation of initially static
  configuration. Top panel: Using the standard scheme, a convection
  like velocity field develops. Shown is absolute velocity (from
  purple to white; sound speed ranging from 0.76 to 1.39) with
  velocity arrows (rainbow colored according to magnitude) for the
  axi-symmetric 256$^{2}$ mesh after $300 \, \tau_{\mathrm{s}}$.
  Bottom panel: Using the well-balanced scheme, an initially static
  polytrope on a 3D Cartesian mesh (128$^{3}$, 'star in a box') is
  preserved to machine precision. Shown are iso-surfaces in
  density after $300 \, \tau_{\mathrm{s}}$. They remain perfectly
    spherical despite the Cartesian grid, because velocities remain
    zero to machine precision, thanks to the well-balanced scheme.
}
\label{LE}
\end{figure}
Lane-Emden polytropes are a good first approximation for the structure
of a star.  They are solutions to the Lane-Emden equation~\citep[see
  e.g.][]{1990sse..book.....K},
\begin{equation}
\label{Eq:Lane-Emden_0}
\frac1{r^2}\frac{d}{dr}\left(r^2\gamma\,\xi\frac{d\rho}{dr}\right)=-4\pi G\rho\,,
\end{equation}
which follows from the equation of hydrostatic equilibrium in 3D spherical symmetry,
$dp / dr = - \rho(r) GM(r) / r^2$,
for a polytropic EoS $p=\xi\rho^{\gamma}$, with $\gamma=1+1/n$ and
$n=1,2,\ldots$ the polytrope index. A few analytical solutions are
known, for instance for $\gamma=2$ ($n=1$), our
test case:
\begin{equation}
\label{eq:init_lane}
\rho(r) = f(\alpha r),  \mbox{\hspace{0.5cm}} p(r) = \xi f(\alpha r)^{2}, \mbox{\hspace{0.5cm}} \phi(r) = -2\xi f(\alpha r).
\end{equation}
Here, $r$ is the radius, $\alpha=\sqrt{2\pi G/\xi\vphantom{A^A}}$, and
$f(\alpha r) = \rho_{c}\,\mathrm{sin}(\alpha r)/(\alpha r)$.  We use
$\xi=1$, $G=1$, and $\rho_c=1$. The situation is marginally stably stratified.

Our goal is again to test whether the well-balanced scheme can
maintain the initial, static situation and how, for comparison, the
standard scheme behaves. We consider two geometries: a 2D
axi-symmetric mesh as well as a 3D Cartesian mesh. 
The discretized analytical solution, Eq.~\ref{eq:init_lane} exactly
satisfies the discrete hydrostatic equilibrium (\ref{Eq_discrete}) for
both meshes, thus can be used as numerical initial condition.

In the 2D axi-symmetric case, we map the computational to
the physical mesh via the mapping function 
$[x,y,z] \rightarrow [r,\theta, \varphi]$, with $[0\ldots1,0\ldots1,
  \Delta z] \longmapsto [0.19\ldots0.95,0.25\pi\ldots0.75\pi, \Delta
  \varphi]$.
This produces a spherical wedge with a width of one cell in
z-direction. We tested that the result does not depend on the choice
of the size of $\Delta \varphi$.  The domain is discretized by a
regular $256^{2}$ mesh.  In radial direction, historical and
reflecting boundary conditions are used at the top and bottom
boundary, respectively (see
Sect.~\ref{Sec:Tests_QuasiStatic_Slab}). Periodic boundary conditions
are used in angular direction.

With the well-balanced scheme, velocities remain zero up to machine
precision over several hundred sound crossing times (not shown). With
the standard scheme velocities start developing early on. After a few
hundred sound crossing times a flow field featuring roll-ups,
updrafts, and downdrafts has formed, as illustrated in Fig.~\ref{LE},
top panel. The apparent symmetry of the flow structure merely
  mirrors the symmetry of the code and the initial conditions. Mach
numbers are around 0.0005 on average but exceed 0.001 in substantial
portions of the flow (green, yellow, and red arrows). This may not
seem dramatic. Recall, however, that sound speeds typically exceed
convective velocities by one or two orders of magnitude for the
stellar situations of interest here (see
Sect.~\ref{Sec:Algorithm_Code}). This translates the velocities in
Fig.~\ref{LE}, which are of purely numerical origin, into a range
between 1\% to 10\% of typical convective velocities. A serious
numerical contamination of the physics to be studied thus seems
possible if the standard scheme is used.

Using a uniform $128^{3}$ 3D Cartesian mesh as physical mesh on a
$[0,1]^{3}$ domain with historical boundary conditions, we repeated
the above exercise. The approach is of interest for simulating entire
stars, where spherical coordinates struggle with (coordinate-)
singularities in the center and along the z-axis. Our approach, often
termed {\it star in a box}, also suffers from a non-optimal mesh - it
is a spherical problem on a physically Cartesian mesh. Nevertheless,
an initially static configuration is maintained to machine precision
by the well-balanced scheme.  Fig.~\ref{LE}, bottom panel, shows
density after $300 \, \tau_{\mathrm{s}}$. Spherical density contours
are perfectly preserved, despite the Cartesian mesh. The standard
scheme (not shown) develops jittery density contours, the spherical
symmetry is broken, and large scale, convection like velocities
develop that reach Mach numbers of about 0.1.
\subsection{Convective layers}
\label{Sec:Tests_Convection}
So far we have looked at stationary situations with zero-velocities
(up to machine-precision), implying that the energy source term
$\rho{\bf u}{\nabla\phi}$ vanishes as well. In this section, we look
at convectively unstable situations that are, nevertheless, in a
globally stable equilibrium. The test case, to be detailed below,
follows~\citet{1984ApJ...282..557H} (H84 in the following).  It
consists of a gravitationally stratified slab in planar geometry with
parameters and boundary conditions such that compressible convection
develops. H84 presented a large test-suite of different 2D cases. We
repeat two tests for comparison. We then go beyond H84 by briefly
addressing the formation of zonal flows in 2D as well as convection in
3D. These additional tests shall demonstrate that our implementation
is fit for 3D and draw attention to a specific 2D flow regime. They
are not meant to be a parameter study.
\subsubsection{Description of test case}
\label{Sec:convective_atmospheres}
We summarize only some key aspects.  A comprehensive description of
the test can be found in H84. The test has two free parameters: the
assumed stratification of the slab, $\chi$, and the Rayleigh number,
$R$.  Quantities are in dimensionless form.  Gravity points along the
negative $y$-axis.  A perfect mono-atomic gas is assumed with gas
constant $R_{gas} = 1$, specific heats at constant pressure and volume
of $c_{p} = 2.5 R_{gas}$ and $c_{v} = 1.5 R_{gas}$, respectively, and with
$\gamma = c_{p} / c_{v}$. The thermal conductivity $K$ and the dynamic
viscosity $\mu$ are constant, with values such that the viscous
diffusion rate is equal to the thermal diffusion rate, i.e., $\sigma =
\mu c_{p} / K = 1$, with $\sigma$ the Prandtl number.  In the absence
of motion, the mean stratification follows a polytrope with
temperature, density, and pressure given by
\begin{equation}
\label{eq:init_h84}
T = y, \mbox{\hspace{1cm}} \rho = y^{m}, \mbox{\hspace{1cm}} p=y^{m+1}.
\end{equation}
As H84, we take for the polytropic index $m=1$.  In the y-direction,
the (dimensionless) slab extent is $d=1$, with the top and bottom
boundaries of the slab at $y_{t}$ and $y_{b}=y_{t}+1$,
respectively. The vertical coordinate $y$ and the desired
stratification $\chi$ are linked via $\chi = \rho(y_{b}) /
\rho(y_{t})$ or $y_{t} = 1/(\chi - 1)$. As $m = g/R_{gas}\beta_{0} -
1$, with $\beta_{0} = dT/dy = 1$ the initial temperature gradient, it
follows that, again in scaled units, $g=2$.

The degree of instability of this configuration can be measured in terms
of the Rayleigh number (see H84)
\begin{equation}
\label{Ra_def}
R(y)=\frac{Q^2(m+1)}{\sigma}\left[1-(m+1)\frac{\gamma-1}{\gamma}\right]y^{2m-1}\,,
\end{equation}
with
\begin{equation}
\label{Q_def}
Q=\frac{(R_{gas}\,|\beta_0|\,d)^{1/2}d}{(K/(\rho_{0}\,c_p))} = \frac{2.5}{K}\,,
\end{equation}
the ratio of the sound travel time to thermal diffusion. The last
equality exploits that $R_{gas} = \beta_{0} = d = 1$, $c_p = 2.5$, and
uses density normalized as in H84, i.e., $\rho_{0} = \rho(y_t) =
1$. For the (convective) instability to set in, one must have $R >
R_{C}$, where $R_{C}$ is the critical Rayleigh number (see Figure 1 in
H84) and $R = R(y_{t}+1/2)$ is the Rayleigh number evaluated at
mid-level. The choice of $R$, or of the factor $f_{R}$ defined via $R
= f_{R} R_{C}$, together with the choice of $\chi$ yielding $y_{t} =
1/(\chi - 1)$, translates into a choice of $K$, via Eqs.~\ref{Ra_def}
and~\ref{Q_def},
\begin{equation}
\label{K_def}
K = \left( 2.5\,(y_{t}+1/2)/R \right)^{1/2},
\end{equation}
thereby also fixing $\mu = K/2.5$, where we used $\sigma = \mu c_{p} / K = 1$.

Boundary conditions at the bottom and top are stress-free for
horizontal velocities, zero velocity in the vertical direction, fixed
temperature at the top boundary ($T = y_{t}$), and fixed temperature
gradient at the bottom boundary ($dT/dy = 1$). The bottom boundary
translates into a steady energy input into the slab in the form of a
radiative flux (see Eq.~\ref{Fr_def} below). Numerical boundary
conditions in the horizontal direction are periodic.

To trigger convection we perturb the initial velocities by at most
$v_{0} = 10^{-4}$. We tested that varying the perturbation has no
effect on the results. Without any triggering perturbations, the
well-balanced scheme maintains the initial profile and convection does
not develop.

Energy transport in the slab can be split into three contributions:
the convective flux $F_{C}$, the kinetic flux $F_{K}$, and the
radiative flux $F_{R}$ (a fourth contribution, the viscous flux, H84
showed to be negligible),
\begin{equation}
\label{Fc_def}
F_C=-\overline{c_p\,\rho  v_y(T-\overline{T})},
\end{equation}
\begin{equation}
\label{Fk_def}
F_K=-\frac{1}{2}\overline{(\rho v_i v_i) v_y},
\end{equation}
\begin{equation}
\label{Fr_def}
F_R=K\frac{\partial \overline{T}}{\partial  y}.
\end{equation}
The over-bar denotes horizontal (normal to the direction of gravity)
averaging. We evaluate Eqs.~\ref{Fc_def} to~\ref{Fr_def} numerically
for diagnostic purposes in each time step.
\begin{figure}[tbp]
\centering
\includegraphics[width=8.0cm]{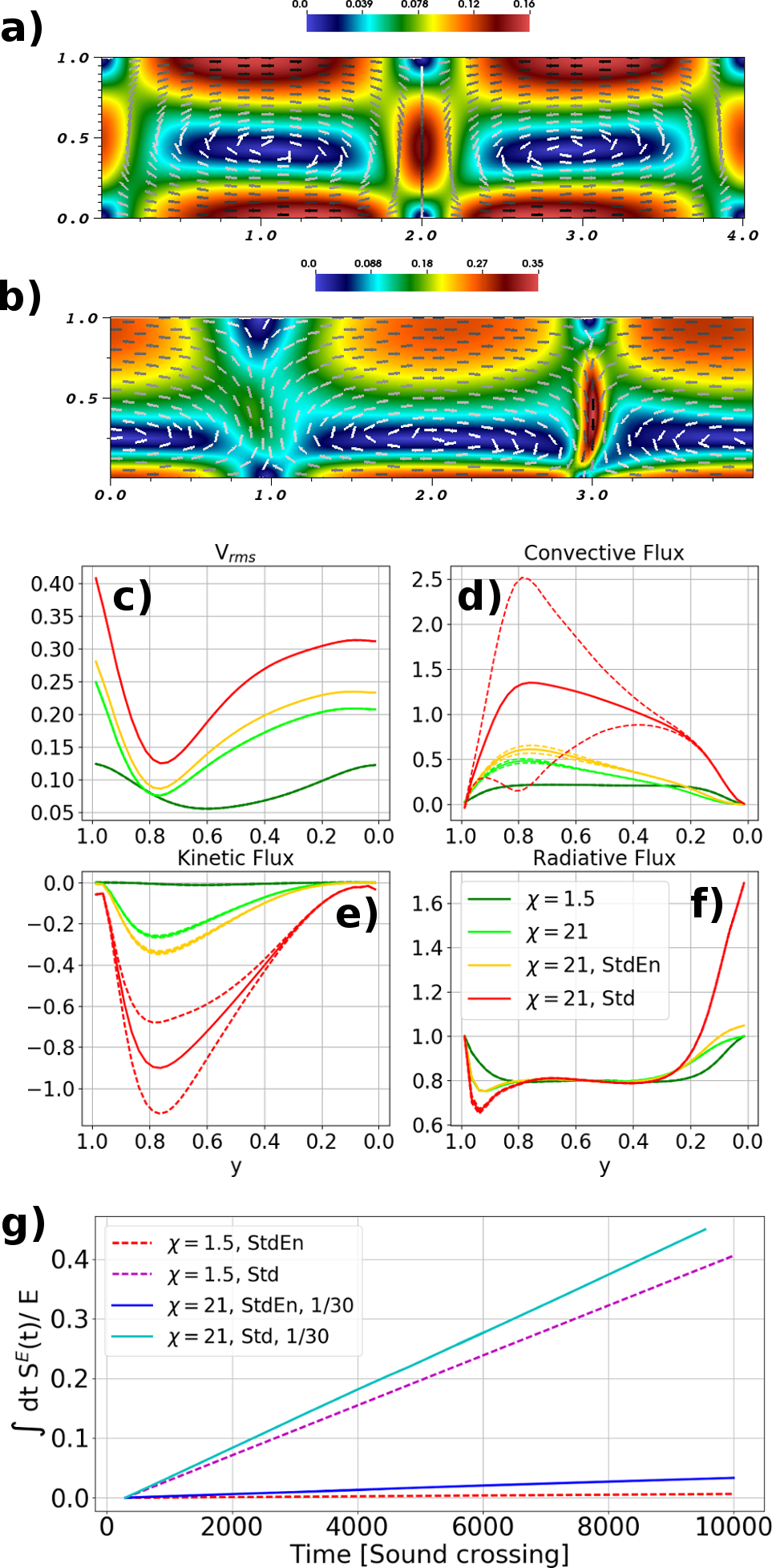}
\caption{H84 test case. Typical flow patterns for $\chi = 1.5$ and
  $\chi = 21$ are shown in panels a) and b), respectively, in (dimensionless)
  absolute velocity (color coded) with velocity arrows superimposed.
  Horizontally averaged vertical profiles of $v_{\mathrm{rms}}$,
  convective ($F_{C}$), kinetic ($F_{K}$), and radiative ($F_{R}$) fluxes
  are given in panels c) to f). Colors denote solutions obtained with
  the well-balanced scheme for $\chi = 1.5$ (dark green) and $\chi =
  21$ (light green), with the well-balanced scheme but central
  differences for $S^{E}$ and $\chi = 21$ (yellow), and with the
  standard scheme for $\chi = 21$ (red). Solid lines are time
  averages, dashed lines are temporal variability ($\pm 1 $ standard
  deviation, where distinguishable from the mean).  Panel g) shows the time integrated gain of energy via
  $S^{E}$ if the scheme is not well balanced in energy (red dashed and
  blue solid) or if the standard scheme (purple dashed and cyan solid)
  is used for $\chi = 1.5$ (dashed lines) and $\chi = 21$ (solid
  lines, scaled by a factor of 30).}
\label{Fig:StableAtmosphere_2D_ConvectiveSlabs}
\end{figure}
\subsubsection{2D simulations}
\label{sec:chi15}
We examine two 2D setups analogous to H84 and show that we obtain
similar results, notably similar convection patterns and energy
fluxes. One setup is mildly stratified ($\chi=1.5$, $R = 310\,R_{C}$,
$R_{C}=400$, $K=7.1 \cdot 10^{-3}$, $\mu=2.8 \cdot 10^{-3}$), the other is more
strongly stratified ($\chi=21$, $R = 1480\,R_{C}$, $R_{C}=750$,
$K=1.1 \cdot 10^{-3}$, $\mu=4.5 \cdot 10^{-4}$), with $R_{C}$ estimated from
Figure 1 in H84.  The computational domain has an extension of 4 in
x-direction, and of 1 in y-direction, with a uniform mesh of $N_x\times
N_y=160\times40$.

The solution obtained with the well-balanced scheme (along with some
not well-balanced solutions, to be discussed below) is illustrated in
Fig.~\ref{Fig:StableAtmosphere_2D_ConvectiveSlabs}. Convective
roll-ups develop that persist as time evolves (panels a and b),
closely resembling results by H84, their Figure 4. Vertical time
averaged profiles of $F_{R}$, $F_{C}$, and $F_{K}$ (green lines in
panels d to f) are equally in line with H84. The radiative flux
accounts for around 80\% of the total flux in both cases. This value
is to be expected for $m=1$, while for $m<1$ the convective flux will
become more important and flow structures will be less
smooth~\citep[see][]{2005AN....326..681B}. In the mildly stratified
case, energy transport via convection accounts for the remaining 20\%
of the total energy flux. In the strongly stratified case, the
convective flux in addition compensates for the downward directed
kinetic flux.  Also shown in
Fig.~\ref{Fig:StableAtmosphere_2D_ConvectiveSlabs} are vertical time
averaged profiles of the root mean square velocity, $v_{rms}$. The
quantity is clearly dominated by the horizontal branches of the
convective motion. As such, $v_{rms}$ mirrors the overall shift of the
convective roll-ups towards the lower boundary of the slab for
$\chi=21$ as compared to $\chi=1.5$. The same dependence was reported
in H84.

To demonstrate the importance of the well-balanced scheme, we repeated
each test case but relaxed the well-balanced properties of the
scheme. The resulting, purely numerical energy gain via the $S^{E}$ is
illustrated in Fig.~\ref{Fig:StableAtmosphere_2D_ConvectiveSlabs},
panel g). The energy gain is found to be more severe for $\chi = 21$
than for $\chi = 1.5$ (by more than a factor of 30) and also more
severe for the standard scheme than for the scheme that is
well-balanced in $S^{M}$ but not in $S^{E}$ (more than a factor of
10). For $\chi = 21$ and the standard scheme, the time integrated
energy gain equals the total gas energy $E$ of the slab after only
about $1000 \, \tau_{s}$. The solution adjusts such that the
temperature gradient steepens close to the top of the slab (not shown)
and the excess energy gained is radiated via $F_{R}$. This excess in
$F_{R}$ near the top boundary is clearly visible in
Fig.~\ref{Fig:StableAtmosphere_2D_ConvectiveSlabs}, panel f, yellow
and red curves, showing the $\chi = 21$ case. In the interior of the
slab, $F_{R}$ remains unchanged whereas $F_{C}$ and $F_{K}$ both
increase, as does $v_{\mathrm{rms}}$.  If the standard scheme is used,
the damage to the solution is particularly severe. Vertical profiles
for $\chi=21$ (red lines in
Fig.~\ref{Fig:StableAtmosphere_2D_ConvectiveSlabs}, panels c to f) now
deviate substantially from the well-balanced solution, with strong
temporal variability in the case of $F_{C}$ and $F_{K}$. The latter is
in line with the 2D flow field no longer being quasi-stationary:
convective plumes move and occasionally the up-draft and down-draft
branch approach so closely that convection collapses and re-forms.
\begin{figure}[tbp]
\centering
\includegraphics[width=8.5cm,height=6.0cm]{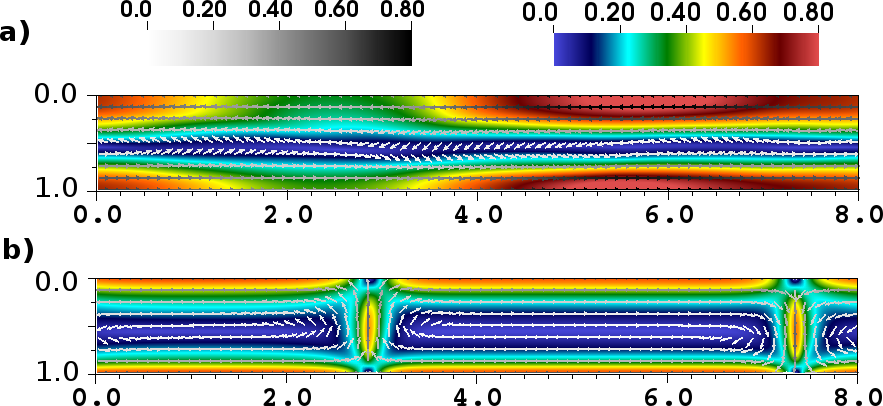}
\caption{H84 2D test case for $\chi = 1.5$ with $\mu = 0$,
  $K=1.4 \cdot 10^{-2}$, and a Rayleigh number of $1.24 \cdot 10^{5}$. The flow
  alters between long zonal-flow phases (panel a) and intermittent,
  convective burst phases (panel b).  Absolute velocity is color
  coded, velocity arrows are overplotted in gray shadings.  }
\label{Fig:H84_ZonalFlow}
\end{figure}
\subsubsection{2D Flows with Low Viscosity or High Rayleigh Number}
We repeated both 2D test cases to highlight two limiting cases that
are well-documented in the literature and are of potential interest in the
context of stellar convection modeling: low viscosity and high
Rayleigh number~\citep[e.g.][]{1995A&A...293..127M,
  2005AN....326..681B, 2014JFM...759..360G, 2014PhRvE..90a3017V}.

Setting $\mu = 0$, we no longer observe quasi-stationary convection
but intermittent, 'bursty' convection. Fig.~\ref{Fig:H84_ZonalFlow},
illustrates the situation for the $\chi = 1.5$ case (with a Rayleigh
number of $1.24 \cdot 10^{5}$, $K=1.4 \cdot 10^{-2}$ and for a domain
size of 8 instead of 4, which, however, has no effect). Strong
horizontal or zonal flows prevail for most of the time (panel a),
interrupted by occasional, short phases dominated by convective
roll-ups (panel b).  A similar behavior is reported
by~\citet{2014PhRvE..90a3017V}.  Going to much higher Rayleigh numbers
while keeping $\sigma = 1$, the convective roll-ups permanently vanish
in favor of zonal flow. A discussion of the underlying mechanisms may
be found in~\citet{2014JFM...759..360G}. Once a strong horizontal flow
component starts developing, it hinders the organization of convection
in the vertical direction. Stress-free boundaries, which are otherwise
well-suited to model part of a stratified
flow~\citep[e.g.][]{1993ApJ...416..733H}, promote the occurrence of
the phenomenon~\citep[e.g.][]{2014JFM...759..360G,
  2014PhRvE..90a3017V}. Note that numerical viscosity is still
  present in all of the above flows, but it is apparently too small to
  prevent the occurrence of zonal flows. While zonal flows are widely
reported in the context of compressible or Rayleigh-B\'{e}nard
convection in 2D or quasi 2D for specific categories of flows - in
particular for flows featuring high Rayleigh number, small viscosity,
and stress-free boundaries - there exist to our knowledge no reports
of zonal flows in truly 3D situations~\citep{1995A&A...293..127M,
  2014JFM...759..360G, 2014PhRvE..90a3017V, 2017PhRvF...2h3501A}.
\subsubsection{3D simulation}
We repeat the $\chi = 1.5$ and $\chi = 21$ setups in 3D to demonstrate
that our algorithm works in 3D and to highlight some differences
between 2D and 3D compressible
convection~\citep[e.g.][]{1995A&A...293..127M, 2000ASSL..254...37L,
  2007IAUS..239..247A, 2009A&A...501..659M, 2015ApJ...815...42G}.  We
use three 3D domains that have identical x- and y- extent as in
Sect.~\ref{sec:chi15} but differ in their z-extent: a 'half bar'
(z-extend 0.5), a 'bar' domain (z-extent 1), and a 'square' domain
(z-extent 4).  Gravity points along the y-axis. Discretization is as
in 2D, i.e., an extent of 1 is covered by 40 cells. We slightly
perturb the velocity in the initialization to trigger convection but
tested that this does not affect the results.

\begin{figure}[tbp]
\centering
\includegraphics[width=9.0cm]{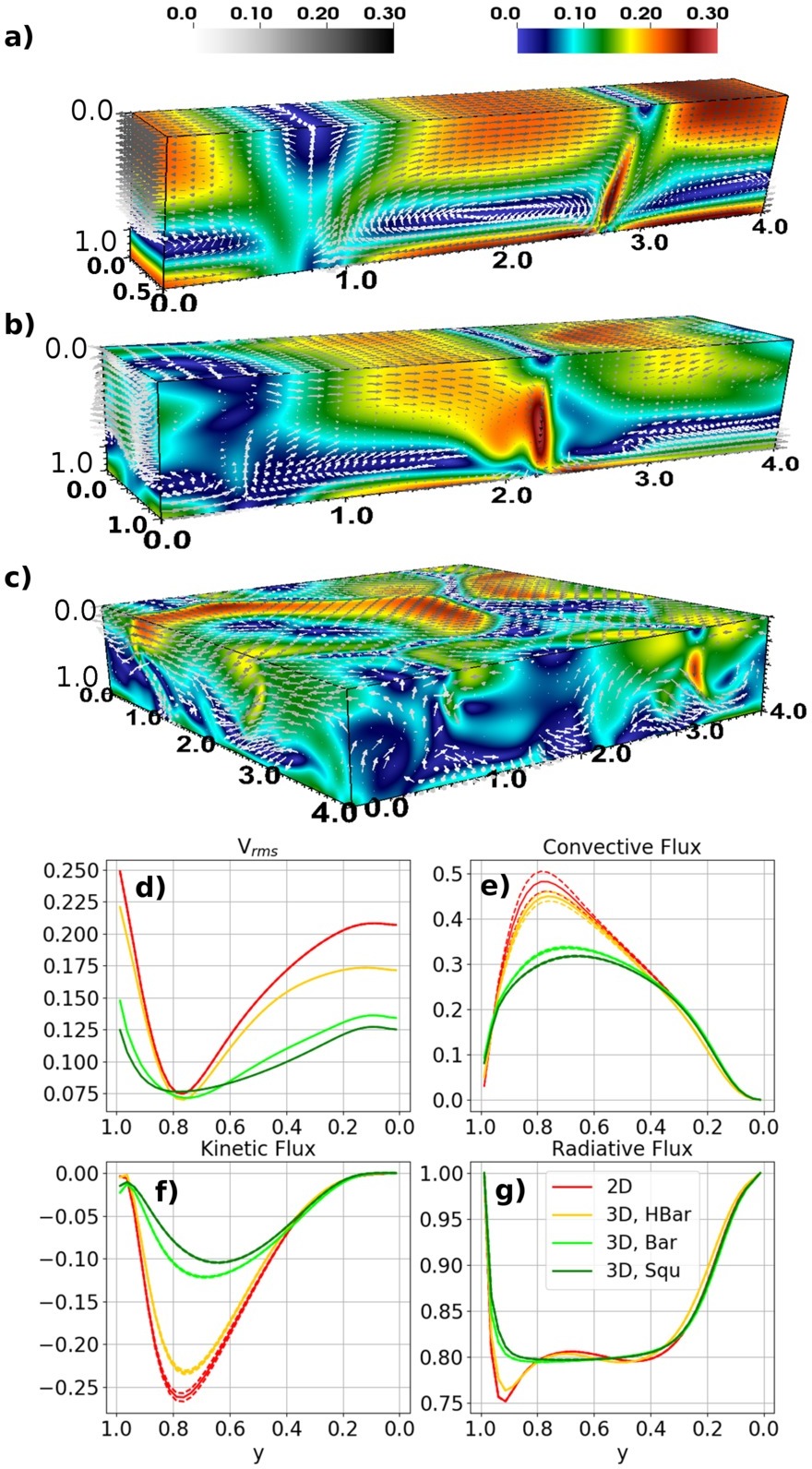}
\caption{Convective H84 slabs in 3D. Shown are the flow field
  (absolute dimensionless velocity color coded) for $\chi = 21$ on
  domains 'half bar', 'bar' and 'square' (panels a to c), as well as
  vertical flux profiles (time averages, solid, and one standard
  deviation, dashed) for $v_{\mathrm{rms}}$, $F_{C}$, $F_{K}$, and
  $F_{R}$ (panels d to g). Discretization is 40 cells per interval $[0,1]$. }
\label{Fig:Hurlbert_3D_8_1_1_Chi=1.5_Non_Visc_Ra=QM}
\end{figure}

Aspects of the settled solutions for $\chi = 21$ are illustrated in
Fig.~\ref{Fig:Hurlbert_3D_8_1_1_Chi=1.5_Non_Visc_Ra=QM}. From the
velocity fields it can be seen that the 'half bar' domain (panel a)
settles into a very similar state to 2D. This organization remains
stable over time. In the 'bar' domain (panel b), the coherence of the
2D flow pattern already tends to be lost. Horizontal velocities near
the top boundary and near the updraft (to the left in the figure) are
no longer aligned with each other. The updraft region occasionally
splits in two. In the 'square' domain (panel c) the solution does not
organize at all. Updrafts and downdrafts form and disappear,
potentially moving around in between. A limitation on the geometrical
degrees of freedom thus seems crucial for the organization of
convection into steady roll-ups, at least in this particular test
case. The relevant energy fluxes in the vertical direction, $F_{R}$
and $F_{C}$, averaged horizontally and over time,
(Fig.~\ref{Fig:Hurlbert_3D_8_1_1_Chi=1.5_Non_Visc_Ra=QM}, panels e to
g) change as well as. In particular, the convective flux $F_{C}$ peaks
at lower values as one goes from 2D over 'half bar' to 'bar' to
'square'. At the same time, the time variability (dashed lines in the
figure) decreases. The root mean square velocity close to the top and
bottom boundary of the slab decreases as the geometrical freedom
increases from 2D to 'half bar', 'bar', and 'square'. Differences are
most pronounced between 2D and 'half bar on the one hand, and 'bar'
and 'square' on the other hand.

The $\chi=1.5$ case shows qualitatively the same behavior, but
differences as one goes from 2D to 'square' domain are generally less
pronounced.
\section{Application to the young sun}
\label{Sec:YoungSun}
As a last test case we move to a more complicated and astrophysically
more relevant situation: the model of a partially convective young
sun. The test aims at demonstrating that the methods introduced work
beyond comparatively simple test cases. In the absence of any
analytical solution of the problem, we benchmark our results against
corresponding, published simulations with the code
MUSIC~\citep{2011A&A...531A..86V, 2013A&A...555A..81V,
  2016A&A...586A.153V, 2017A&A...600A...7G}. Our physical setup, to be
detailed below, follows the one used in these studies. The section
does not aim at a detailed physical comparison of the solutions
obtained by MUSIC and A-MaZe, which is beyond the scope of the present
paper and a topic of future studies.
\subsection{Definition of the problem}
\label{Sec:YoungSun_Defs}
We consider a model of a convective young sun with mass
$M_{star}=1\,M_{\odot}$ and radius $R_{star}=3\,R_{\odot}$ at an age of
a few Myr.  We use the the same realistic stellar, tabulated EoS
$p=p(\rho,e)$ (or, equivalently, $T=T(\rho,e)$) as
in~\citet{2016A&A...593A.121P} (see references therein).  Inversion of
the EoS is done numerically. Heat transfer is non-linear and the heat
conduction $K(T)$ is, within the diffusion approximation for radiative
transfer, given by the approximation
\begin{equation}
\label{Rflux}
K(T) = \frac{16\sigma T^3}{3\kappa\rho}\,, 
\end{equation}
where $\sigma$ is the Stefan-Boltzmann constant and $\kappa$ is the
tabulated Rosseland opacity of the gas \citep{1996ApJ...464..943I,
  2005ApJ...623..585F}. The explicit viscosity $\mu$ is assumed to be
zero, as the actual physical value is much smaller than the numerical
viscosity. We use the same 1D initial model as used
in~\citet{2016A&A...593A.121P}.  The gravitational potential $\phi$ is
taken from the 1D simulations and is kept fixed in time. The
  average Mach number of the problem ranges from around 0.001 in the
  bulk of the convective zone to about 0.01 close to
  surface~\citep[see also][]{2016A&A...586A.153V}.

We perform five simulations that differ in their resolution and
geometry / dimensionality. Four simulations use an axi-symmetric
geometry, on grids of $64^2$, $128^2$, $256^2$ and $512^2$. One
simulation uses a 3D computational mesh of $64^3$. A uniform mapping
as in Sect.~\ref{Sec:Tests_QuasiStatic_Polytrope_AxiSym} is used:
$(x,y)\rightarrow(r,\theta)$ in the 2D axi-symmetric case and
$(x,y,z)\rightarrow(r,\theta,\varphi)$ in the 3D case with uniform
grid spacing. In the radial direction and in units of stellar radius
$R_{star}$, the computational domain extends from $R_{min}=0.21$ to
$R_{max}=0.94$ ('Low 3' case in~\citet{2016A&A...593A.121P}), thereby
comprising both, the convective envelope and parts of the radiative
core. The domain extends from $0.2\pi$ to $0.8\pi$ in polar angle
$\theta$ (and in azimuth angle $\varphi$ in 3D).

Boundary conditions are periodic in angular directions. The
  slight dependence of the ghost cell volume for meridional directions
  (between a few percent and fractions of a percent, depending on
  resolution) is neglected, yet the fluxes at the periodic boundaries
  respect the conservative scheme. In radial direction, we set the
mass fluxes at the domain boundary to zero (stress free for tangential
velocities, reflecting for radial velocities), retain only the
pressure term for the momentum fluxes (to ascertain well-balance), and
prescribe fixed (radiative) energy fluxes taken from the initial 1D
profile: $1.28 \cdot 10^{32}$ ergs/s at the inner boundary and $8.91
\cdot 10^{33}$ ergs/s at the outer boundary, which are then both
scaled according to the fraction of the full sphere contained in the
simulation domain. The luminosity $L$ at the top and bottom of the
domain is taken from the initial 1D profile.  Note that more energy is
radiated at the outer boundary than enters through the inner boundary,
in line with the young sun being slowly contracting.  To sustain a
reasonable temperature profile close to the top boundary, despite our
rather coarse numerical grid, we apply Newtonian
cooling~\citep{2006ApJ...638..336D, 2011A&A...531A..86V}.
\begin{figure}[tbp]
\centering
\includegraphics[width=8.2cm]{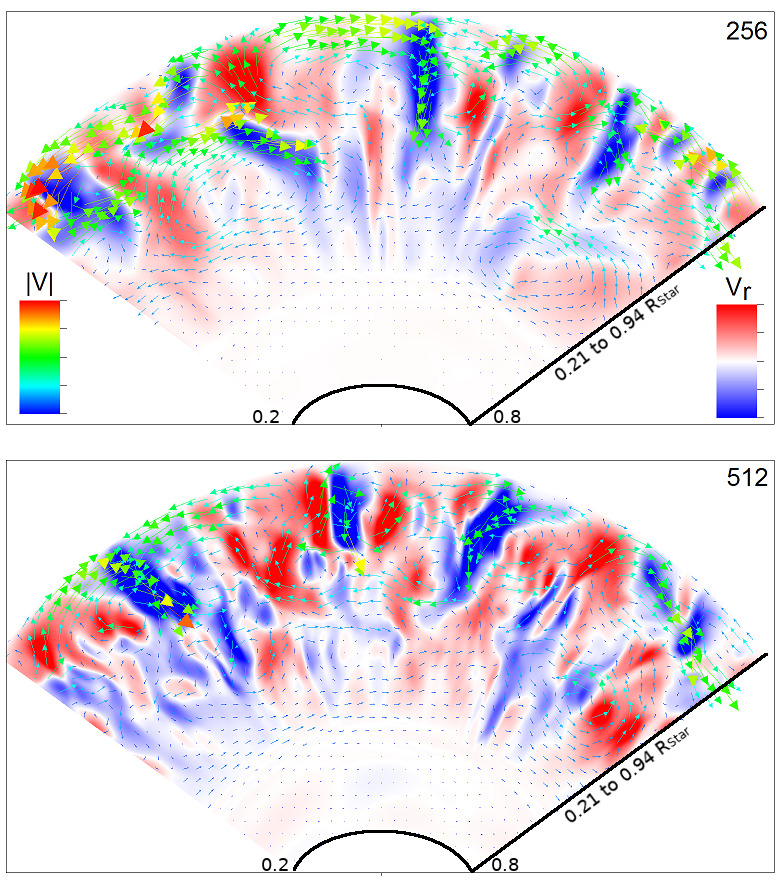}
\caption{Quasi-stationary convective structure of the young sun on the
  mesh $256^2$ (top panel) and $512^{2}$ (bottom panel). Shown is
  radial velocity (color coded, from blue, at -3000 m/s downward,
    to red, at +3000 m/s upward) with velocity arrows (rainbow colors
  according to magnitude, linear from 0 to 10$^{4}$m/s) after
  about $10^{7}$ seconds (for comparison with
  Fig.~\ref{Fig:LogEkin_Logt}).  Color coding is the same in both
  panels. The domain extends in radial direction from 0.21 to 0.94 in units of $R_{star}$,
  and from 0.2$\pi$ to 0.8$\pi$ in meridional direction.  }
\label{Fig:StellarConvection1}
\end{figure}
\begin{figure}[tbp]
\centering
\includegraphics[width=9.0cm]{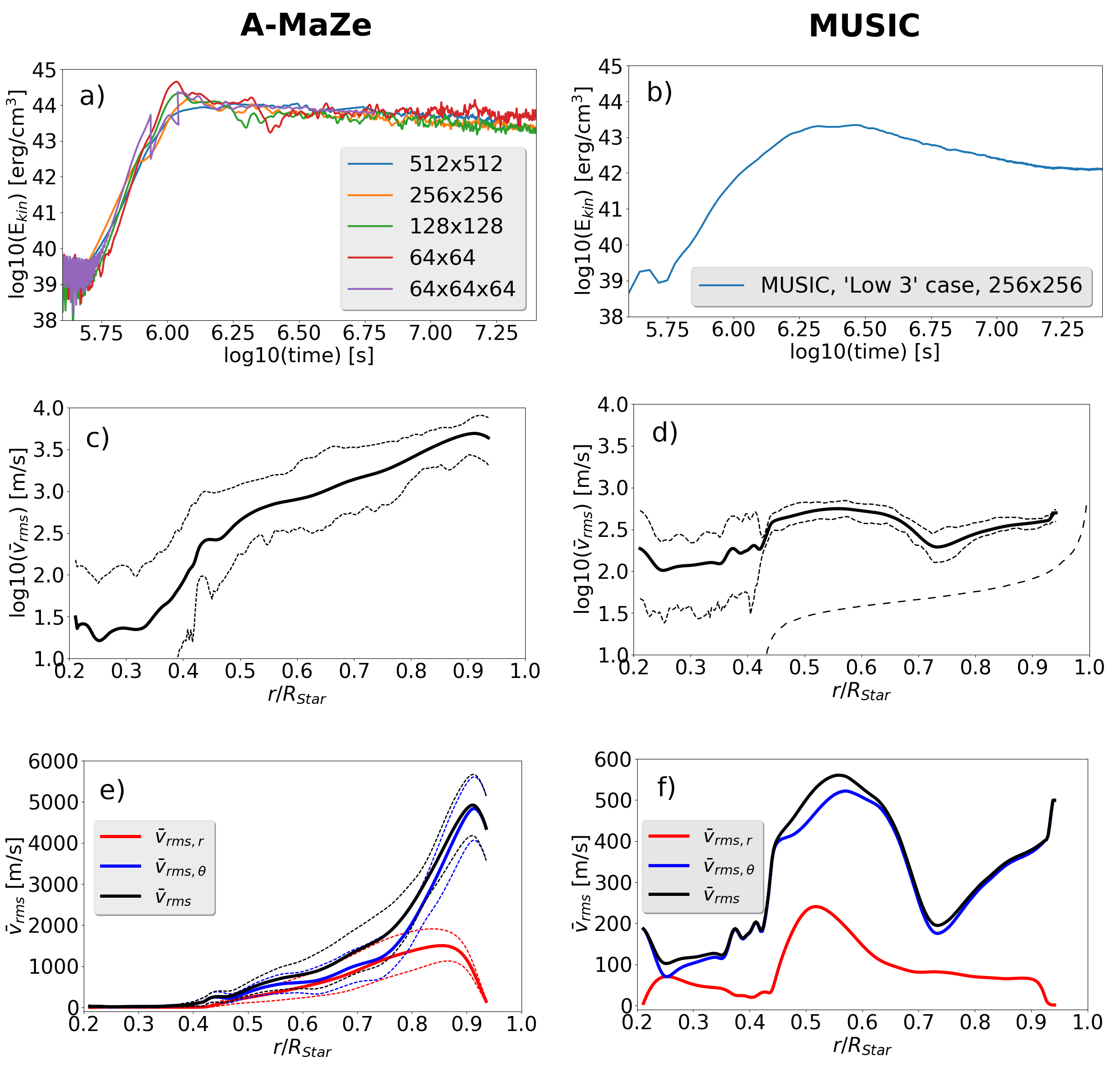}
\caption{Axi-symmetric young sun simulations, 2D, with A-MaZe (left
  panels, $256^{2}$ mesh unless otherwise stated) and MUSIC (right panels, 'Low 3'
  case in~\citet{2016A&A...593A.121P}). Shown are the total kinetic
  energy in the domain, scaled to a full spherical layer
  ($0\le\theta\le\pi$, $0\le\varphi\le2\pi$, $0.21\le R \le 0.94$), as
  function of time (different resolutions color coded for A-MaZe,
  shifted in time such as to ramp up simultaneously), as well as
  radial profiles of volume-weighted $v_{rms}$ and its decomposition
  into radial and tangential component (middle and bottom panels,
  solid lines give time averages, dotted lines indicate the range)
  after convection has become quasi-stationary. The black long-dashed line
  (middle right panel) shows the convective velocity calculated within 
  Mixing Length Theory formalism. Note the different y-axes in panels e) and f).}
\label{Fig:LogEkin_Logt}
\end{figure}
\subsection{Axi-symmetric simulations}
\label{Sec:YoungSun_axi-symmetric}
An illustration of the fully developed convection in the form of 2D
velocity maps is given in Fig.~\ref{Fig:StellarConvection1} for the
resolutions $256^{2}$ and $512^{2}$, comparable to cases 'Low' and
'Hi' in~\citet{2016A&A...593A.121P}. The separation of the radiative core from the
convective envelope is clearly apparent (transition to predominantly
white color), at the same radial distance for both resolutions
shown. Structures appear overall finer in the $512^{2}$ resolution
case than in the $256^{2}$ case, as expected. The difference
is particularly apparent when looking at the up- and down-drafts
(colored in red and blue). Velocities tend to reach higher peak values
in the $512^{2}$ case and are generally higher closer to the upper
boundary.

The bottom panel of Fig.~\ref{Fig:StellarConvection1} roughly
corresponds to Figure~5, left panel, in~\citet{2016A&A...593A.121P}.
Comparing the two figures, they indeed look similar. In both figures,
the separation between the radiative core and the convective envelope
is located at about 1/3 of the radius shown. The up- and down-drafts
appear rather more fine-grained and slightly more elongated in
Fig.~\ref{Fig:StellarConvection1} than in \citet{2016A&A...593A.121P}.

Convection in the A-MaZe simulations is overall more vigorous than in
the MUSIC simulations, in terms of total kinetic energy or also with
regard to $v_{rms}$, both shown in Fig.~\ref{Fig:LogEkin_Logt}. The
total kinetic energy in the A-MaZe simulations is comparable for all
resolutions, but systematically exceeds the value in MUSIC by roughly
an order of magnitude.  Turning from the total kinetic energy to the
radial dependence of $v_{\mathrm{rms}}$, some more facets emerge. In
the convection zone, $v_{\mathrm{rms}}$ is larger in A-MaZe than in
MUSIC (Fig.~\ref{Fig:LogEkin_Logt}, middle panels). More specifically,
both codes produce a comparable $v_{\mathrm{rms}}$ close to the
boundary between the radiative and the convective zone. But while in
MUSIC there is little net increase of $v_{\mathrm{rms}}$ with radius
within the convective zone, $v_{\mathrm{rms}}$ increases by about one
order of magnitude in A-MaZe. Out to about $0.8\,R_{\mathrm{star}}$,
the radial and tangential component of $v_{\mathrm{rms}}$ increase in
concert, reaching values somewhat above $\sim1000$~m/s
(Fig.~\ref{Fig:LogEkin_Logt}, bottom panels). Meanwhile, in MUSIC
$v_{rms}$ reaches only $\sim 250$~m/s, yet again tangential velocities
dominate for large radii. The difference also impacts the convective
turnover time, which is often used as a characteristic time
scale:
the overall larger radial velocities in A-MaZe result in an overall
shorter convective turnover time compared to MUSIC.  At yet larger
radial scales, beyond $0.8\,R_{\mathrm{star}}$, a large part of
$v_{rms}$ in A-MaZe is contained in tangential velocities along the
stellar surface.  In the radiative zone, by contrast, A-MaZe yields
robustly lower $v_{\mathrm{rms}}$ than MUSIC
(Fig.~\ref{Fig:LogEkin_Logt}, middle panels).

In summary, the above findings suggest qualitative agreement between
A-MaZe and MUSIC, enhancing overall confidence in the results obtained
by both codes. Yet quantitative differences exists. A detailed
analysis of the underlying causes is beyond the scope of this
paper. The analysis performed with MUSIC
in~\citet{2016A&A...593A.121P} highlights the sensitivity of
convective velocities to boundary conditions, extension of the radial
domain and resolution, easily yielding a factor 5 difference between
results based on different setups. Algorithmic differences may also
contribute. On a more general level, the low Mach number limit of
  the compressible Euler (or Navier-Stokes) equations remains
  challenging, despite much progress in terms of physical
  understanding and numerical
  handling~\citep[e.g][]{guillard:hal-00871725, 2010JCoPh.229..978D,
    guillard:hal-01534938, 2019JCoPh.393..278A}. From this perspective, the differences
documented here provide a basis for further numerical and physical
progress.
\begin{figure}[tbp]
\centering
\vspace{1cm}
\includegraphics[width=8.7cm]{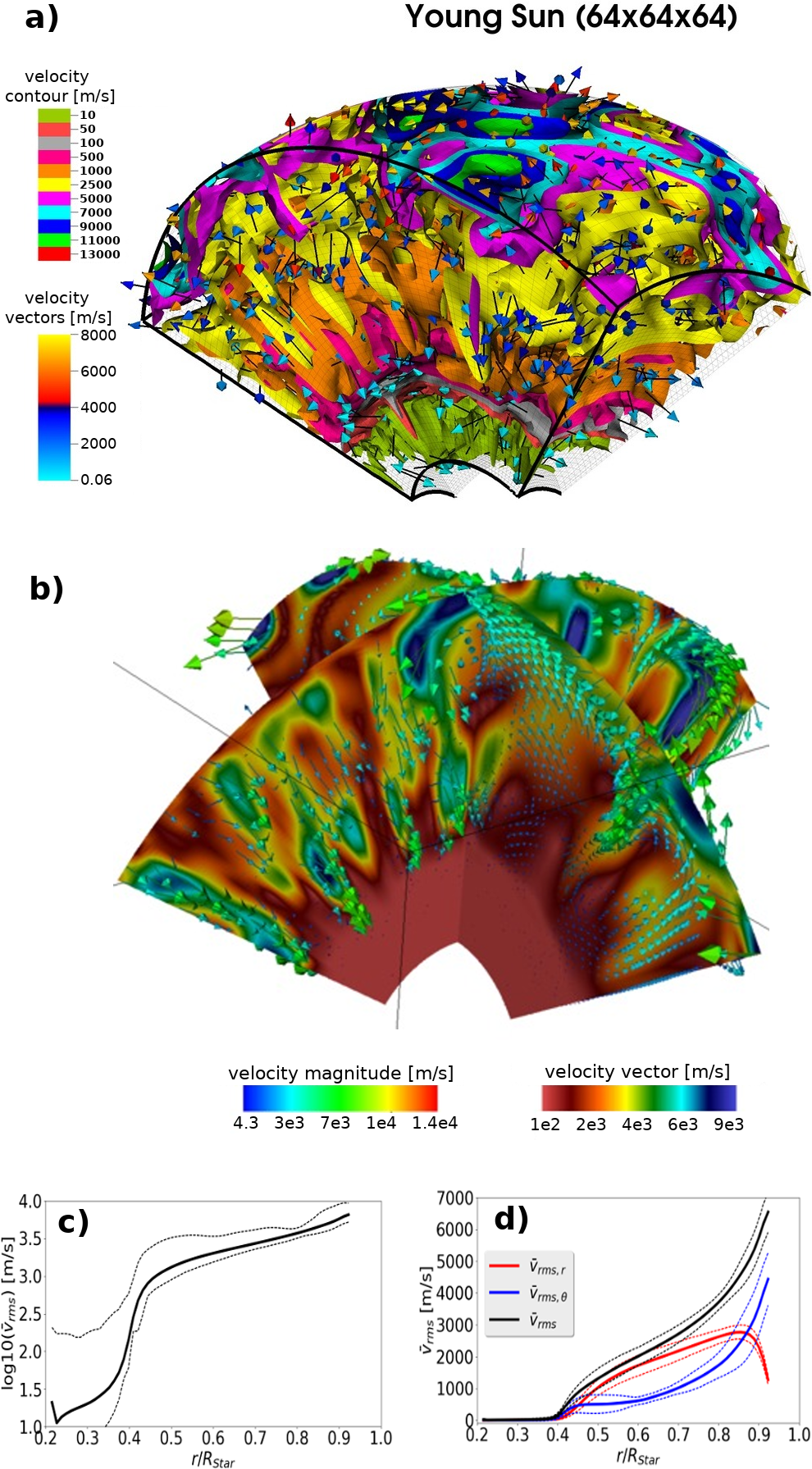}
\caption{Young sun simulation in 3D, $64^{3}$ mesh. Panel a: Surfaces
  of constant velocity along with velocity arrows. Panel b: slices
  along the equator and perpendicular to the equator. Shown is
  velocity magnitude (color coded) along with velocity arrows (arrows
  color coded according to velocity magnitude). Bottom panels: radial
  profiles of $v_{rms}$ (panel c, logarithmic scale) and of $v_{rms}$
  split into radial and tangential components (panel d, total
  $v_{rms}$ also shown). Solid lines show time averages, the range
  (minimum to maximum) is indicted by dotted lines.}
\label{Fig:3D_StellarStructure_surfaces}
\end{figure}
\subsection{3D simulations}
\label{Sec:YoungSun_3D}
The results from the 3D simulation are illustrated in
Fig.~\ref{Fig:3D_StellarStructure_surfaces}, at a time when convection
has become quasi-stationary. The goal of these simulations is not the
physical study of the young sun. The grid of $64^3$ is too coarse
to reasonably resolve numerous relevant physical structures, such as
the strong temperature and density gradients characterising the
near-surface layers. The purpose is rather to illustrate that the
well-balanced scheme can cope with the application, despite coarse
resolution and more complicated physics, notably in terms of EoS.

Pockets of high-velocity material are visible in the top panel of
Fig.~\ref{Fig:3D_StellarStructure_surfaces}, in cyan, blue, and green
colors. The separation of the convective and the radiative zone is
visible in Fig.~\ref{Fig:3D_StellarStructure_surfaces}, middle panel,
as the transition to uniformly reddish colors. Also clearly visible in
the same panel are numerous up- and down-drafts, some of the latter
penetrating into the radiative zone.  Across the transition from the
convective to the radiative zone, $v_{rms}$ drops by a factor between
ten to hundred, as can be taken from the bottom panels in
Fig.~\ref{Fig:3D_StellarStructure_surfaces}. Radial profiles of
$v_{rms}$ are qualitatively similar to their 2D counterparts (see
Fig.~\ref{Fig:LogEkin_Logt}). Also similar is the dominance of the
tangential component for $v_{rms}$ for radii larger than about $0.8 \,
R_{star}$, roughly the same radius as in the 2D simulations. The
kinetic energy of the 3D $64^{3}$ simulation is comparable to that in
the $256^{2}$ and $512^{2}$ simulations in 2D
(Fig.~\ref{Fig:LogEkin_Logt}, top left panel).

We take the above as an indication that our well-balanced
implementation works equally well in both, 2D and 3D.
\section{Discussion}
\label{Sec:Discussion}
In the following, we want to put the well-balanced extension to A-MaZe
somewhat more into context. We will argue at the same time that the
well-balanced extension - despite or because of its simplicity - makes
A-MaZe a valuable tool for the numerical investigation of
multidimensional stellar convection, especially in concert with other
codes.

We noted already in Sect.~\ref{Sec:Introduction} that there exist
various approaches to cope with a gravitational field, be it external
or due to self-gravity. With our choice of algorithm and its concrete
implementation, we compromised on accuracy (second order in space in
our implementation) and generality (stationary potential) in favor of
simplicity and efficiency, both with regard to coding and execution
time. The algorithm is compact and self-contained so that it can be
implemented within an existing scheme or framework without
difficulty. Additionally, the extension of the scheme to higher order
accuracy for the non-hydrostatic part of the problem ($p^{1}$ in
Sect.~\ref{SubSec:WB_MomentumBalance}) is straightforward. Going
beyond second order for the hydrostatic reconstruction ($p^{0}$ in
Sect.~\ref{SubSec:WB_MomentumBalance}) is more complex. The underlying
assumption of $\phi$ being piece-wise linear is exactly fulfilled in
the tests of Sect.~\ref{Sec:Tests}, but only up to truncation errors
in the case of the young sun, Sect.~\ref{Sec:YoungSun}. In real
application cases, the solution thus is potentially vulnerable again
to too coarse meshes. However, this may not be a too severe
restriction. In many astrophysical applications, $\phi$ varies more
slowly and smoothly as a function of space and time than other physics
of interest. A computational mesh fit for the latter is thus likely
fine enough to somewhat mitigate the second order only reconstruction
of $p^{0}$ in practical applications.  Issues potentially also exist
in multi-dimensions if the gravitational force is not aligned with a
coordinate axis, although practical applications do not appear to be
severely impacted by this (see KM16 for a discussion). For the tests
and applications of interest here, the restriction to a time-constant
gravitational potential is not an issue. We see no principle obstacle
as to why our approach should not be extendable to cases where $\phi$
is time dependent. In fact, KM16 present in their paper a
  corresponding test case, a toy model of a core-collapse supernova.
It then would be interesting to compare the source term based approach
presented here and a pure flux formulation as
e.g. in~\citet{2013NewA...19...48J}.

Further advantages of the scheme exist beyond the previously mentioned
simplicity and efficacy.  The approach used here does not make any
assumption on the thermal equilibrium, does not rely on a fixed mesh
size, and allows for any time integration scheme, explicit or implicit
(see KM16).  Also, our cell-centered finite-volume scheme lends itself
more easily to adaptive meshes than do staggered grids. The latter is
potentially of interest in the context of our envisaged applications
to stellar convection. In such applications, it may be desirable to
have higher resolution at large stellar radii, towards the stellar
surface, or also in the vicinity of ionization edges further
inward. Finally, we have demonstrated in Sect.~\ref{Sec:YoungSun} that
the current implementation, without adaptive grids, is useful for
studying stellar convection: sufficiently long, physical integration
times are possible despite explicit time stepping, and results are
qualitatively comparable to MUSIC results. Associated integration
  times cannot be compared as the simulations were run on different
  machines. Comparisons of integration times for MUSIC alone may be
  found in~\citet{2013A&A...555A..81V} (explicit versus previous
  implicit solver) and~\citet{2016A&A...586A.153V} (previous versus
  current implicit solver).

Multidimensional studies of the interior dynamics of stars, from core
to atmosphere, are still not common place and pose multiple,
non-trivial challenges for numerical simulations. To ascertain the
physical robustness of simulation results it is then advantageous to
study the same situation with different simulation codes. This
overarching idea resulted in the two codes
MUSIC~\citep{2011A&A...531A..86V, 2013A&A...555A..81V,
  2016A&A...586A.153V} and the well-balanced hydro code within the
A-MaZe tool-kit presented here. Similarities between both codes are,
briefly, that they simulate hydrodynamic convection in stellar
interiors in 2D and 3D for a realistic EoS. Major differences between
A-MaZe and MUSIC concern the grid (cell-centered versus staggered) or
the time integration (explicit versus implicit).

The differences between the two codes translate into different
code-dependent strengths and weaknesses. The present paper
demonstrates that both codes are basically fit to simulate stellar
convection, thereby opening the possibility to duplicate at least some
simulations or part thereof with both codes.  Arriving at similar
physical results with both codes greatly strengthens their
credibility.  Likewise, qualitative or quantitative difference point
the way to where further improvement of numerics and / or physics is
needed.
\section{Summary and conclusions}
\label{Sec:Conclusion}
We equipped the multi-scale, multi-physics numerical tool-kit A-MaZe
with a well-balanced algorithm that balances both, momentum and energy
of a flow in the presence of a static gravitational field.
The balancing with regard to energy is less of a topic in literature
than the momentum balance, at least if the equations are formulated
with source terms and not as a conservation law for the total energy
of the gas, i.e., internal plus kinetic plus gravitational energy.

A series of tests are presented that demonstrate the capabilities of
our implementation, even at low resolution. An initially static
configuration can be maintained to machine precision, even for a
spherical setup on a Cartesian mesh. A quasi-stationary convective
situation can be simulated without net gain or loss of energy despite
strong local up-drafts and down-drafts. If a not (fully) well-balanced
scheme is used, a substantial and steady energy gain on purely
numerical grounds is shown to occur that affects the solution. The
convection test is further used to exemplify differences between 2D
and 3D convection, including the occurrence of zonal flows in 2D
convection.  Application of our code to a young sun in 2D and 3D,
comprising an inner radiative and an outer convective part, completes
our tests. Simulation results show reasonable agreement with published
results.

With A-MaZe and MUSIC we now have two largely independent codes within
our collaboration, with which to study multidimensional stellar
convection or also simpler setups of compressible convection, like the
ones used here primarily for code testing. Being able to repeat
selected simulations or parts thereof with different codes promotes
credibility of the obtained numerical results in case they are
reasonably similar, or points the way to necessary physical and / or
numerical improvements in case of substantial disagreement. The
present work may also provide a direction of travel to defining
benchmark cases for stellar convection - instead of usual
Rayleigh-B\'{e}nard convection - that would be useful for the rest of
the community interested in the study of compressible convection.
\begin{acknowledgements}
  We acknowledge support from the European Research Council through
  grant ERC-AdG No. 320478-TOFU. RW and DF acknowledge support from
  the French National Program for High Energies PNHE.  RW and DF are
  also grateful for the hospitality and fruitful discussion with the
  colleagues, in particular Professor Keh-Ming Shyue, of the Center
  for Theoretical Sciences, Mathematical Division at National Taiwan
  University where part of this work has been made. Dr. Jean M. Favre
  from the Swiss Supercompute Center (CSCS) we thank for sharing his
  profound knowledge on visualization and data analysis. An anonymous
  referee we thank for most constructive comments. Simulations have
  been performed at the P\^{o}le Scientifique de Mod\'{e}lisation
  Num\'{e}rique (PSMN), and at centers of the Grand Equipement
  National de Calcul Intensif (GENCI) under grant number
  A0310406960. We acknowledge the steady support of the staff at these
  two centers.

\end{acknowledgements}
%
%
%
%
%
%
%
\bibliographystyle{aa} 
\bibliography{WellBalanced} 
\end{document}